# Stochastic interference in a dispersive nonlinear optical fiber system

**Shiva Kumar and Jing Shao**[*]

*Department of Electrical and Computer Engneering, McMaster University, Hamilton ON, L8S 4L8, Canada*
[*]*jingshao.hust@gmail.com*

**Abstract:** Stochastic power fluctuation in a fiber optic system due to the interplay among dispersion, nonlinearity and partial coherence of the source is investigated. An analytical expression for the power fluctuation of a signal pulse due to its interference with an echo pulse generated due to the nonlinear interaction of signal pulses in a fiber optic system excited by a partially coherent source is obtained. The analytical results show that the mean nonlinear distortion decreases as the coherence time of the source reduces consistent with our numerical simulations.

## 1. Introduction

As an optical pulse propagates in a fiber, it broadens due to dispersion [1,2]. When the fiber is excited by a partially coherent source, the pulse undergoes additional broadening and the amount of broaderning depends on the degree of coherence of the source. Marcuse derived an expression for the pulse broadening in optical fiber taking into account the partical coherence property of the source and showed that the amount of pulse broadening increases significantly if the source spectral width is much larger than the signal bandwidth [2]. Lin et al introduced the concept of partially coherent Gaussian Schell-Model pulse in time domain to describe the pulse propagation in dispersive media [3]. Recently, we generalized the results of Marcuse [2] to multiple optical pulses with the introduction of correlation functions that describe stochastic interference of pulses and developed an analytic expression for the correlation functions [4]. When the source is coherent, pulses broaden and overlap with other pulses and interference fringes (in time domain) occur over a long transmission distance. However, as the coherence time of the source decreases, visibility of fringes drops after a short distance. In the analysis of [4], we had ignored the fiber nonlinear effects. In this paper, we extend the analysis of [4] to include the dynamic interplay between dispersion, nonlinearity and partial coherence of the source. When nonlinearity is taken into account, correlation function describing the interference between the signal pulse and echo pulse needs to be calculated. Picozzi [5] et al have obtained different forms of kinetic equations to accurately describe the evolution of the correlation function of the partially coherent wave propagating through the nonlinear medium.

In a high bit rate, highly dispersive single channel system, signal pulses broaden significantly and interact nonlinearly with other pulses leading to intra-channel four wave mixing (IFWM) [6-12] and intra-channel cross phase modulation (IXPM) [12-14]. In the absence of dispersion, the temporally separated pulses would not overlap and hence, IFWM and IXPM would be absent. Three signal pulses centered at $T_1$, $T_2$ and $T_3$ interact nonlinearly in a highly dispersive fiber system leading to echo pulses at $T_1 + T_2 - T_3$, $T_1 + T_3 - T_2$, and $T_2 + T_3 - T_1$ due to IFWM. Mecozzi et al developed an analytical expression for the evolution of the echo pulse using a first order perturbation theory [8-9], and it is assumed that the fiber is excited by a coherent source. In this paper, we analyze the nonlinear interaction of signal pulses in a highly dispersive fiber system excited by a partially coherent source and develop an analytical expression for the evolution of the echo pulse, when the source correlation function and signal pulse shape are both Gaussian. On-off keying modulation is used in this paper for simulation and discussion, and the results may not only be limited in on-off keying system, but also multi-level amplitude-shift keying system. Our analytical results show that as the coherence time of the source decreases, the nonlinear distortion reduces. The numerical simulation of a long haul fiber optic system excited by a partially coherent source also shows the same trend. The reasons for the reduction in nonlinear distortion can be explained as follows: in the case of fully coherent source, the signal pulses in various bit slots are in phase (unless there is a phase modulation) and the efficiency of nonlinear mixing of the signal pulses to generate echo pulses increases due to this phase matching. In contrast, in the case of partially coherent source, the signal pulses in adjacent bit slots could have randomly varying phases depending on the degree of coherence. In addition, the visibility of interference fringes (in time domain) due to the interference of echo pulses with signal pulses drops as the coherence time decreases, which implies lower power fluctuations. Our numerical simulation result shows that the nonlinear distortion is not a stationary random process for the given bit pattern due to the time dependence caused by the bit pattern.

In this paper, we compare two types of dispersion maps: (i) dispersion managed (DM) system and (ii) dispersion unmanaged (DU) system. In DM systems, the dispersion of each transmission fiber is compensated for using an inline dispersion compensating fiber (DCF). In DU systems, there is no inline DCF and dispersion of transmission fibers is compensated for using an DCF at the end of the fiber optic link or it is compensated in electrical domain. Our results show that for DU systems, as the spectral width of the source changes from 2 kHz (coherence time = 1 ms) to 2 GHz (coherence time = 1 ns), the mean nonlinear distortion drops by 14% whereas for DM systems, it hardly changes. This result can be explained as follows. For DU systems, the pulses broaden a lot over the long haul fiber optic link and a signal pulse centered at $t$=0 s interacts nonlinearly with signal pulses located upto 150 bit slots (i.e. upto $\pm 6$ ns) on either side. The coherence time of 1 ns implies that pulses within a period of ~1 ns are strongly correlated and hence the nonlinear interaction of the pulse centered at $t$ =0 s with the pulses that are located beyond $\pm 1$ ns is reduced. However, for the DM systems, the pulses do not broaden a lot and a signal pulse centered at $t$ =0 s interacts nonlinearly with pulses located upto 15 bitslots (i.e. upto $\pm 600$ ps). Since the coherence time of 1 ns is larger than this nonlinear interaction time, there is no reduction in nonlinear distortion as compared to the nearly coherent case of $\tau_0 = 1$ ms.

The rest of the paper is organized as follows. Section 2 reviews the stochastic power fluctuations in a dispersive fiber when the nonlinear effects are ignored. Section 3 generalizes the results of Section 2 by including the fiber nonlinearity. The analytical and numerical results are shown in Section 4.

## 2. Stochastic interference in a dispersive linear system

In this section, we will review the stochastic interference in a dispersive linear system, which has been published in [4]. Suppose the source of the fiber optic communication system is not an ideal laser, but a partically coherent source. Its output is

$$\psi_{in}(t) = A(t)e^{-i\omega_0 t}, \tag{1}$$

where $\omega_0$ is the mean angular frequency and $A(t)$ is a stationary random process with the correlation function

$$R(\tau) = \langle A(t)A^*(t+\tau) \rangle, \tag{2}$$

where $\langle \ \rangle$ denotes the ensemble average. Assuming a Gaussian spectral distribution of the source with the spectral width $W$, the source correlation function is

$$R(\tau) = P_{in} e^{-\frac{\tau^2}{\tau_0^2}}, \tag{3}$$

where $P_{in}$ is the input power and $\tau_0 = 2/W$ is the coherence time.

Let us first consider the case of a single pulse launched to the fiber. The field at the fiber input is

$$\psi(t,0) = q(t,0)e^{-i\omega_0 t} \tag{4}$$

$$q(t,0) = A(t)p(t), \tag{5}$$

$$p(t) = e^{-\frac{t^2}{2T_0^2}}, \tag{6}$$

where $q(t,0)$ is the input field envelope, $p(t)$ is the pulse shape function, and Guassian shape is used. In Eq. (6), $T_0$ is the half-width at 1/e-intensity point. The signal propagation in a linear dispersive fiber is described by [1]

$$i\frac{\partial q}{\partial z} - \frac{\beta_2}{2}\frac{\partial^2 q}{\partial t^2} = 0, \tag{7}$$

where $\beta_2$ is the second order dispersion coefficient. After propagating through a fiber of length $L$, the mean output optical power is

$$P_0(t) \equiv P(t,L) = \langle |q(t,L)|^2 \rangle. \tag{8}$$

Marcuse has derived an analytical expression for the stochastic broadening as

$$P_0(t) = \frac{P_{in}}{\eta} e^{-\frac{t^2}{(\eta T_0)^2}}, \quad (9)$$

where

$$\eta = \left[1 + \left(\frac{\beta_2 L}{T_0^2}\right)^2 + \left(\frac{2\beta_2 L}{T_0 \tau_0}\right)^2\right]^{1/2}. \quad (10)$$

For a coherent source, $\tau_0 = \infty$ and the third term in Eq. (10) vanishes. For a partially coherent source, from Eqs. (9) and (10), we see that the pulse undergoes additional broadening due to the interaction of dispersion and partial coherence (third term in Eq. (10)).

Next, let us consider the case of two pulses launched to the fiber. The input field envelope is

$$q(t,0) = A(t)\left[a_1 p\left(t + \frac{T_s}{2}\right) + a_2 p\left(t - \frac{T_s}{2}\right)\right], \quad (11)$$

where $a_1$, $a_2$ are the transmission data, $T_s$ is the temporal separation. We have developed an analytical expression for the mean output power as [4]

$$P(t,L) = |a_1|^2 P_0\left(t + \frac{T_s}{2}\right) + |a_2|^2 P_0\left(t - \frac{T_s}{2}\right) + a_1 a_2^* \Gamma(t,T_s) + a_1^* a_2 \Gamma^*(t,T_s), \quad (12)$$

where $P_0(t)$ is given by Eq. (9). The function $\Gamma(t,T_s)$ is similar to the self-coherence function [15-17]. $\Gamma$ accounts for the stochastic interference of the fields due to temporally separated pulses. An analytical expression for $|\Gamma(t,T_s)|$ is derived in [4] which is given by

$$|\Gamma(t,T_s)| = \frac{P_{in}}{\eta} e^{-\frac{1}{(\eta T_0)^2}\left[t^2 + \left(\frac{T_s \lambda}{2}\right)^2\right]}, \quad (13)$$

where

$$\lambda = \sqrt{1 + \frac{4\beta_2^2 L^2}{\tau_0^2 T_0^2}}. \quad (14)$$

As the coherence time $\tau_0$ decreases, $\lambda$ increases and $|\Gamma(t,T_s)|$ decreases. For a fully incoherent source, $\tau_0 = 0$ and $|\Gamma(t,T_s)| = 0$; in this case, the output power is simply the addition of powers due to individual pulses.

Finally, we consider the case of a random bit pattern launched to the fiber. In this case, the input field envelope is

$$q(t,0) = A(t) \sum_{n=-\infty}^{\infty} a_n p(t - nT_s), \quad (15)$$

where $\{a_n\}$ is the random bit pattern that takes values 0 and 1 with equal probability and assumed to be real. The mean output power is [4]

$$P(t,L) = \sum_m \sum_n a_m a_n^* \Gamma_{m,n}(t,T_s), \quad (16)$$

where

$$\Gamma_{m,n}(t,T_s) = \frac{P_{in}}{\eta} e^{-\frac{T_0^2 t^2 + \sigma t + \xi}{(\eta T_0^2)}}, \quad (17)$$

$$\sigma = -(m+n+1)T_s(T_0^2 + iS) + 2iT_s S(n+1/2), \quad (18)$$

$$\xi = -\frac{1}{2}i(m+n+1)(n-m)T_s^2 S + (n-m)^2 \frac{T_s^2 S^2}{\tau_0^2} + \frac{1}{2}\left(m^2 + n^2 + m + n + \frac{1}{2}\right)T_s^2 T_0^2, \quad (19)$$

and

$$S = \beta_2 L, \quad (20)$$

is the accumulated dispersion.

## 3. Stochastic interference in a dispersive nonlinear system

Nonlinear Shrödinger equation (NLSE) is used to describe the pulse propagation in fibers, which is given by

$$i\frac{\partial u}{\partial z} - \frac{\beta_2(z)}{2}\frac{\partial^2 u}{\partial t^2} = -\gamma e^{-\alpha z}|u|^2 u, \qquad (21)$$

where $u = qe^{-\alpha z/2}$, and $q$ is the complex field envelope, $\alpha$, $\beta_2$ and $\gamma$ are loss, dispersion and nonlinear coefficient, respectively. In quasi-linear systems, the leading order solution is obtained by setting $\gamma = 0$ and nonlinearity is treated as a small perturbation on the linear field. The leading order solution is [1]

$$u^{(0)}(t,z) = \sum_{m=-\infty}^{\infty} a_m u_m(t,z), \qquad (22)$$

where $u_m(t,z)$ is the linear solution. We assume that the signal pulses are Gaussian,

$$u_m(t,z) = A(t)\frac{T_0}{T_1}e^{-\frac{(t-mT_s)^2}{2T_1^2}}, \qquad (23)$$

$$T_1^2(t,z) = T_0^2 - iS, \qquad (24)$$

where $A(t)$, $T_0$ and $a_m$ are defined in Section 2 and $S$ is the accumulated dispersion $S(z) = \int_0^z \beta_2(s)ds$.

The field $u(t,z)$ may be expanded in a perturbation series, [18]

$$u(t,z) = u^{(0)}(t,z) + \gamma u^{(1)}(t,z) + \gamma^2 u^{(2)}(t,z) + ..., \qquad (25)$$

where $u^{(m)}$ is the $m$th order solution. Substituting Eq. (25) into Eq. (21), and collecting all the terms that are proportional to $\gamma$, we find the governing equation for the first order solution as

$$i\frac{\partial u^{(1)}}{\partial z} - \frac{\beta_2}{2}\frac{\partial^2 u^{(1)}}{\partial t^2} = -\gamma e^{-\alpha z}\left|u^{(0)}\right|^2 u^{(0)} = -\gamma e^{-\alpha z}\sum_j\sum_k\sum_l u_j u_k u_l^*. \qquad (26)$$

The right hand side (RHS) of Eq. (26) is a summation of multiplication of three pulses centered at $jT_s$, $kT_s$ and $lT_s$. If we focus on one term of Eq. (26), it can be written as

$$i\frac{\partial \varepsilon}{\partial z} - \frac{\beta_2}{2}\frac{\partial^2 \varepsilon}{\partial t^2} = -\gamma e^{-\alpha z} u_j u_k u_l^*, \qquad (27)$$

where $\varepsilon$ is the first order echo pulse resulting from the nonlinear interaction between pulses centered at $t = jT_s$, $kT_s$ and $lT_s$. The solution of Eq. (27) is (see Appendix A)

$$\varepsilon(t,z) = \frac{iT_0^3 \gamma a_j a_k a_l}{2\pi}\int_{-\infty}^{+\infty}\int_0^z ds d\omega e^{-\alpha s + i\beta_2 \omega^2 (z-s)/2 + i\omega t}\int_{-\infty}^{+\infty}dt'\frac{1}{|T_1(s)|^2 T_1(s)}|A(t')|^2 A(t')e^{-\Lambda t'^2 + Bt' + C}e^{-i\omega t'}. \qquad (28)$$

When the source is coherent ($\tau_0 = \infty$), $A(t)$ is a constant and for this case, Mecozzi et al obtained the first order solution as [8]

$$\varepsilon(t,z) = i2\pi T_0^3 \gamma P_{in}^{3/2}\int_0^z ds \frac{1}{|T_1(s)|^2 T_1(s)\sqrt{1+2i\beta_2 s\Lambda}}e^{-\alpha s + C + \frac{B^2}{4\Lambda} - \frac{(B-2\Lambda t)^2}{4\Lambda(1+2i\beta_2 s\Lambda)}}, \qquad (29)$$

where $\Lambda$, $B$, and $C$ are given in Appendix A. The average power at the fiber output is

$$P(t, L_{tot}) = \left\langle \left| u^{(0)}(t, L_{tot}) + \varepsilon(t, L_{tot}) \right|^2 \right\rangle \cong \left\langle \left| u^{(0)}(t, L_{tot}) \right|^2 \right\rangle + 2 \left\langle \mathrm{Re}\left\{ u^{(0)*} \varepsilon(t, L_{tot}) \right\} \right\rangle \quad (30)$$

$$= P_L(t) + P_{NL}(t),$$

where $L_{tot}$ is the total transmission distance. In Eq. (30), we assume that the nonlinearity is a small purtubation on the linear pulse and hence, we ignore the second order term $\left\langle \left| \varepsilon(t, L_{tot}) \right|^2 \right\rangle$.

The first term on the RHS of Eq. (30) is the same as that derived in Section 2 (see Eq. (16)), which takes into account the interplay between dispersion and partial coherence. The second term, $P_{NL}(t)$ represents the nonlinear distortion due to the dynamic interplay between dispersion, nonlinearity and partial coherence. To remove the impact due to the interplay between dispersion and partial coherence of the type discussed in Section 2, dispersion compensating fiber (DCF) is used to fully compensate for the dispersion, so that the field at the end of the fiber optic link is the same as that at the input in the absence of nonlinearity, i.e.,

$$P_L(t) = \left\langle \left| u^{(0)}(t, 0) \right|^2 \right\rangle. \quad (31)$$

Under this condition, there is no stochastic broadening due to the interplay between dispersion and partial coherence at the fiber output. We consider a dispersion managed (DM) fiber system in which the transmission fiber is a positive dispersion fiber whose dispersion is compensated by an inline DCF as well as a dispersion unmanaged (DU) fiber system in which the DCF is introduced at the end of the transmission link. Without loss of generality, we consider the power fluctuation due to nonlinearity in the bit slot 0 centered at $t = 0$.

$$P_{NL}(t) = 2 \left\langle \mathrm{Re}\left\{ u_0^*(t, 0) \varepsilon(t, L_{tot}) \right\} \right\rangle. \quad (32)$$

Note that we have replaced $u^{(0)}$ of Eq. (30) with $u_0$. This is justified due to the fact that the tails of the pulses centered in the neighboring bit slots around $t=0$ are negligibly small. Using Eqs. (23) and (28), Eq. (32) becomes

$$P_{NL}(t) = \frac{T_0^3 \gamma a_j a_k a_l a_0}{\pi} e^{-\frac{t^2}{2T_0^2}} \mathrm{Re}\left\{ \left\langle iA(t)^* \int_0^{L_{tot}} ds e^{-\alpha s} \frac{1}{\left| T_1(s) \right|^2 T_1(s)} \right. \right.$$

$$\left. \left. \times \iint \left| A(t') \right|^2 A(t') e^{-\Lambda t'^2 + Bt'^2 + C} e^{-i\omega t'} e^{i\beta_2 \omega^2 z/2} e^{i\omega t} e^{-i\beta_2 \omega^2 s/2} d\omega dt' \right\rangle \right\},$$

$$= \frac{T_0^3 \gamma a_j a_k a_l a_0}{\pi} e^{-\frac{t^2}{2T_0^2}} \mathrm{Re}\left\{ i \int_0^{L_{tot}} ds e^{-\alpha s} \frac{1}{\left| T_1(s) \right|^2 T_1(s)} \right. \quad (33)$$

$$\left. \times \iint \left\langle A(t)^* \left| A(t') \right|^2 A(t') \right\rangle e^{-\Lambda t'^2 + Bt'^2 + C} e^{-i\omega t'} e^{i\beta_2 \omega^2 z/2} e^{i\omega t} e^{-i\beta_2 \omega^2 s/2} d\omega dt' \right\},$$

In order to simplify Eq. (33) further, we need to evaluate the correlation function appearing inside the double integral. Using Eq. (2), it can be shown that (see Appendix B)

$$R_{NL}(\tau) = \left\langle A(t) \left| A(t) \right|^2 A^*(t+\tau) \right\rangle$$

$$= \frac{1}{2\pi} P_{in}^2 e^{-\frac{\tau^2}{\tau_0^2}}. \quad (34)$$

Since the echo pulse is generated due to the terms of the form $A(t)\left|A(t)\right|^2$, $R_{NL}(\tau)$ may be interpreted as the correlation function describing the interference between the signal pulse $u_0$ and the echo pulse components. Using Eq. (34) in Eq. (33), it can be simplified as

$$P_{NL}(t) = 8\pi^3 a_0 a_j a_k a_l \gamma T_0^3 P_{in}^2 e^{-\frac{t^2}{2T_0^2}} \text{Re}\left\{i\int_0^{L_{tot}} ds\, e^{-\alpha s} M(s) e^{Y(s)}\right\}, \tag{35}$$

where

$$M(s) = \frac{2\tau_0}{|T_1(s)|^2 T_1(s)\sqrt{\tau_0^2 + i2\beta_2 s(1+\Lambda\tau_0^2)}}, \tag{36}$$

and

$$Y(s) = \frac{i\beta_2 s\left(4tB\tau_0^2 + B^2\tau_0^4 - 4t^2\tau_0^2\Lambda\right) - 2t^2\tau_0^4\Lambda + 2tB\tau_0^4}{2\tau_0^2\left[\tau_0^2 + 2i\beta_2 s(1+\Lambda\tau_0^2)\right]} + C. \tag{37}$$

Equation (35) is the main contribution of this paper. It describes the stochastic power fluctuation of the pulse centered at $t=0$ due to its interference with the echo pulse resulting from the nonlinear interaction between pulses centered at $jT_s$, $kT_s$ and $lT_s$. Total power fluctuations of the pulse centered at $t=0$ can be obtained by summing over all the triplets $\{jkl\}$ with the condition $j+k-l=0$. Also, the stochastic power fluctuation of the pulse centered at $t=hT_s$ can be obtained in a similar way by introducing the condition $j+k-l=h$.

## 4. Results and Discussion

Figure 1 shows the schematic of a dispersion-managed (DM) fiber optic system. The following parameters are assumed throughout the paper unless otehwise specified: symbol period, $T_s = 40$ ps, full width at half-maximum (FWHM) of the signal pulses = 8 ps, peak launch power = 2.5 mW. The dispersion, loss and nonlinear coefficients of the standard single mode fiber (SSMF) are -21 ps²/km, 0.2 dB/km, 1.1 km⁻¹W⁻¹, respectively. The length of the SSMF is 80 km per span. The dispersion, loss and nonlinear coefficients of the DCF are 140 ps²/km, 0.4 dB/km, 4.4 km⁻¹W⁻¹, respectively and the length of inline DCF is 12.6 km. The inline amplifiers compensate for the loss of fibers. The noise introduced by the amplifiers is ignored since the main focus of this paper is to study the interplay between dispersion, nonlinearity and partial coherence. The total transmission distance ($L_{tot}$) is 800 km. Let us first consider the case of an optical fiber excited by a coherent source ($\tau_0=\infty$). Two pulses separated by $T_s$ interact nonlinearly to generate echo pulses on both sids of the signal pules, as shown in Fig. 2, which is obtained by solving Eq. (21) numerically using a split-step Fourier scheme [1][12]. If the signal pulses at $T_s$ and $2T_s$ are $u_1(t)$ and $u_2(t)$, respectively, the echo pulses at $t=0$ and $3T_s$ are generated due to the terms $u_1^2 u_2^*$ and $u_2^2 u_1^*$ (see Eq. (26)), respectively.

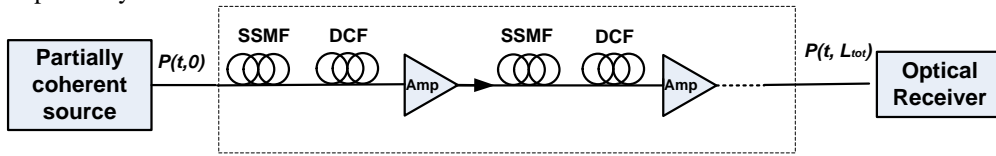

Fig.1 Schematic of a dispersion-managed fiber optic system.

If there is a signal pulse at $t=0$, the interference of the echo pulse with the signal pulse leads to the power fluctuations of the signal pulse. The power fluctuations of the pulse in the bit slot 0 (centered at $t=0$) is not only due to the echo pulse shown in Fig. 2, but also due to SPM ($|u_0|^2 u_0$) and IXPM terms ($|u_j|^2 u_0$, $j=2, 3$) in Eq. (26).

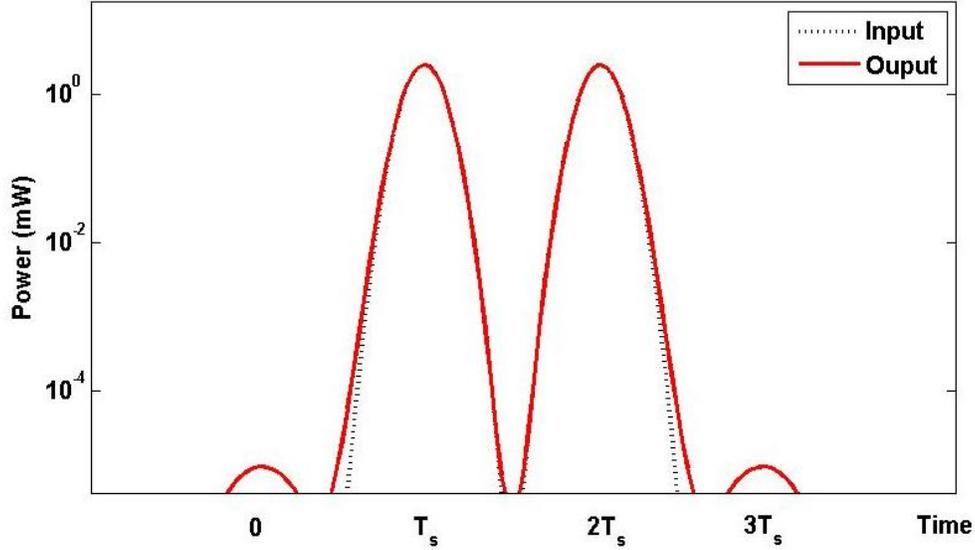

Fig.2 Input and output powers as a function of time in a dispersion-managed fiber optic system. Peak launch power=2.5 mW.

So far we assumed that the source is coherent. Next, we consider the power fluctuation of the signal pulse centered at $t = 0$ when the source is partially coherent. Figure 3 shows the mean power change of the signal pulse, $P_{NL}(t)$ due to its interference with the echo pulse generated by the nonlinear interaction of pulses centered at $T_s$ and $2T_s$ (due to the term. $u_1^2 u_2^*$, i.e. $j = k =1, l = 2$ ), calculated using Eq. (35). On-off keying modulation is introduced throughout this paper unless otherwise specified. When the coherence time $\tau_0$ is much larger than the signal pulse width, the source may be considered as coherent (see solid line in Fig. 3). As can be seen, the mean nonlinear distortion reduces as the coherence time decreases.

In our theoretical analysis, we considered a single echo pulse interfering with a signal pulse centered at $t = 0$. However, in practice, there could be many ( $> 100$) echo pulses interfering with the signal pulse in a strongly pulse overlapped system. In order to estimate the power fluctuations caused by the fiber nonlinearity, we solved the NLSE numerically. A random bit pattern consisting of 16 bits ($a_n$=0 or 1) is used as the fiber input. 1000 independent runs of NLSE solver is carried out using the same bit pattern to calculate the ensemble average. Since the accumulated dispersion at the fiber output is zero, the power fluctuation $P_{NL}(t)$ is simply the difference between the powers at the fiber output and input, averaged over 1000 runs. Figure 4 shows the ensemble-averaged nonlinear distortion $P_{NL}(t)$ as a function of time calculated numerically for different coherence times. As can be seen the power fluctuations reduce as the coherence time decreases consistent with the analytical results shown in Fig. 3. The reasons for the reduction in power fluctuations are twofold: first, the efficiency of echo pulse generation is reduced when the fiber is excited by a partially coherent source (due to the phase mismatch between signal pulses) as compared to the case of a fully coherent source; second, the visibility of interference fringes (in time domain) due to the interference of echo pulses with signal pulses drops as the coherence time decreases. $P_{NL}(t)$ may be interpreted as the

signal-echo pulse beating noise and its variance is

$$\sigma_{NL}^2 = \frac{1}{T_w} \int_{-T_W/2}^{T_W/2} [P_{NL}(t)]^2 dt - (\bar{P}_{NL})^2, \tag{38}$$

$$\bar{P}_{NL} = \frac{1}{T_w} \int_{-T_W/2}^{T_W/2} P_{NL}(t) dt$$

where $T_w$ is the width of the computational window in time domain. Table 1 shows that the standard deviation of nonlinear distortion is reduced as the coherence time decreases. In long-haul direct detection fiber-optic systems, nonlinear distortion is one of the dominant impairments and Fig. 4 (and Table 1) shows that it can be reduced by reducing the coherence time of the source.

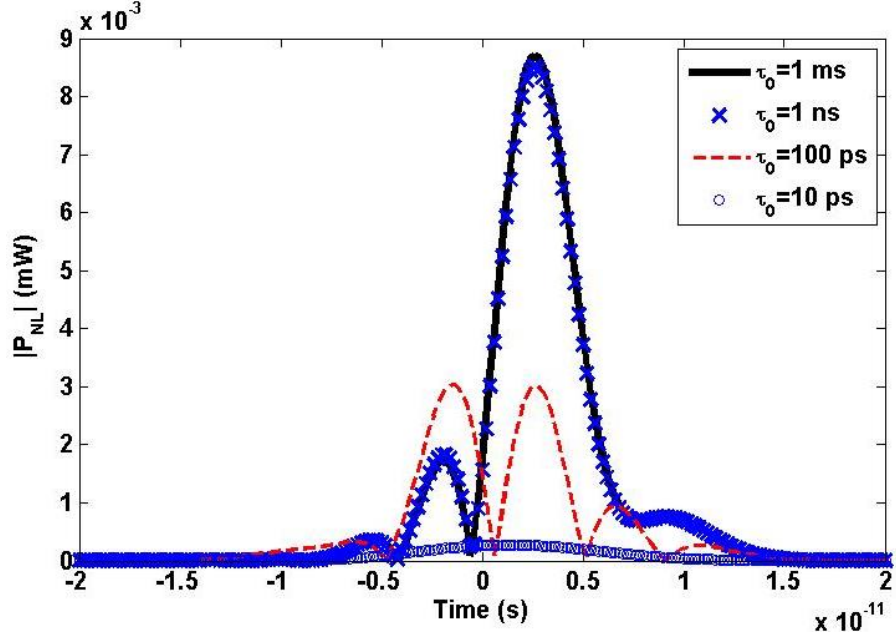

Fig.3. Nonlinear distortion calculated analytically using Eq. (35) in DM system.

From Fig. 4, we see that the nonlinear distortion is not a stationary random process since the fluctuations are time-dependent. It is likely to approach a stationary process if the dispersion is very large, which could erase the bit-pattern dependent (i.e. time dependent) power fluctuations. In a long haul dispersion unmanaged (DU) system, the dispersion is compensated in electrical domain or using DCF. In a DU system, the nonlinear distortion is treated as noise with Gausian distribution even when the fiber is excited by a coherent source [19-21]. In contrast, in the DM systems, the dispersion is compensated in each span and hence, the pulses with large temporal separation ($>15T_s$) would not interact nonlinearly, which leads to bit-pattern dependent power fluctuations.

Next we consider a dispersion unmanaged (DU) system shown in Fig. 5. The fiber parameters, launch power and total transmission distance of the DU fiber system are the same as that of the DM system. The length of DCF is so chosen as to compeletely compensate for the dispersion of the SSMFs. Figure 6 shows the mean power change of the signal pulse due to its interference with the echo pulse generated by the nonlinear interaction of pulses centered at $T_s$ and $2T_s$ in the DU system. In Fig. 6, we find that as the coherence time changed from 1 ms (spectral width = 2 kHz) to 1 ns (spectral width = 2 GHz), the peak drops by 14% for DU system indicating that the nonlinear penalty can be reduced by increasing the spectral width of the source for the commonly employed sources such as DFB lasers. In contrast, for

DM systems, as can be seen from Fig. 3, there is a little change as the coherence time changed from 1 ms to 1 ns. This can be explained as follows. In the case of DU system, pulses broaden a lot and a pulse centered at $t=0$ s interacts nonlinearly with pulses located upto 150 bitslots (i.e. upto $\pm 6$ ns) on either side. If the pulses in these bitslots are strongly correlated (as in the case of fully coherent source), nonlinear penalty is enhanced. The coherence time of 1 ns implies that pulses within a period of ~1 ns are strongly correlated and hence the nonlinear interaction of the pulse centered at $t=0$ s with the pulses that are located beyond $\pm 1$ ns is reduced. However, for the DM systems, the pulses do not broaden a lot and a pulse centered at $t=0$ s interacts nonlinearly with pulses located upto 15 bitslots (i.e. upto $\pm 600$ ps). Since the coherence time of 1 ns is larger than this nonlinear interaction time, there is hardly any reduction in nonlinear distortion as compared to the nearly coherent case of $\tau_0 = 1$ ms.

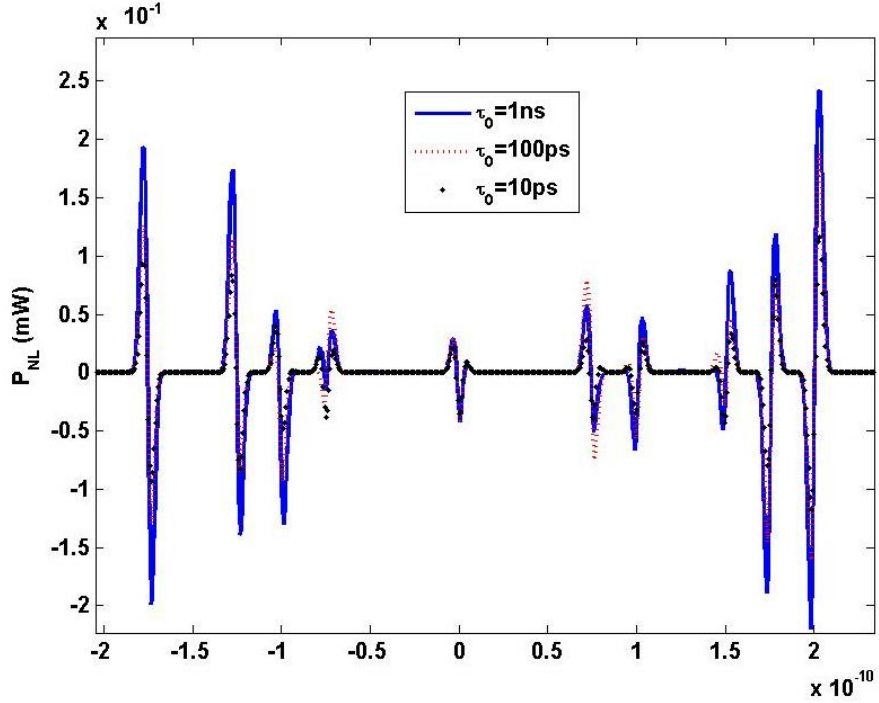

Fig.4. Ensemble-averaged nonlinear distortion as a function of time for various coherence times in DM system.

Table 1: Mean nonlinear distortion for the DM system. $T_w$=640 ps.

| Coherence time $\tau_0$ (ns) | $\sigma_{NL}$ (mW) |
|---|---|
| 1 | 0.0386 |
| 0.1 | 0.0281 |
| 0.01 | 0.0189 |

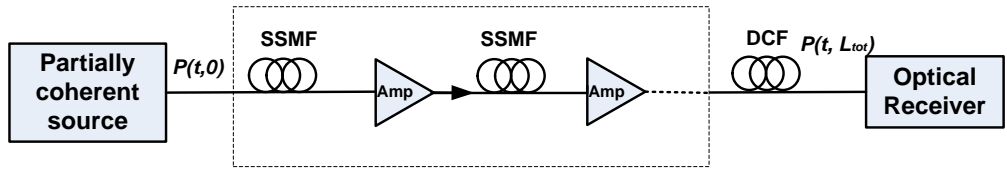

Fig.5 Schematic of a dispersion-unmanaged fiber optic system.

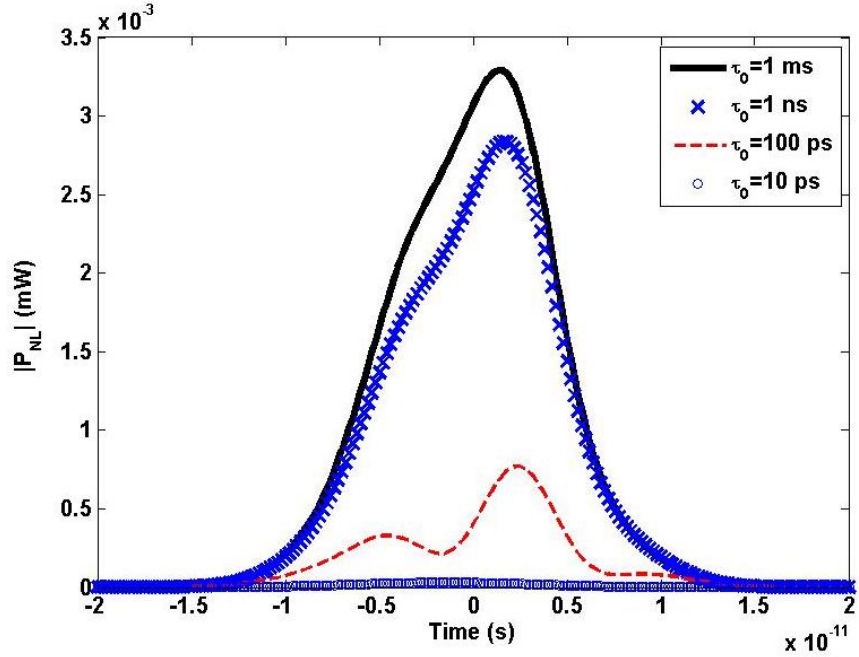

Fig.6. Nonlinear distortion calculated analytically using Eq. (35) for DU system.

Figure 7 shows the ensemble-averaged nonlinear distortion as a function of time calculated numerically for different coherence times in DU fiber system. Comparing Figs. 4 and 7, we find that the peak of the nonlinear distortion in each bit slot is sometimes higher for the case of DU system. However, the negative peak of the nonlinear distortion found in Fig. 4 for the DM systems is reduced or absent for the DU systems. Table 2 shows the standard deviation of nonlinear distortion for the DU system and comparing it with Table 1, we see that the standard deviation is significantly lower for the DU systems as compared to the DM systems for the given coherence time.

## 5. Conclusion

We have investigated the stochastic power fluctuations in a dispersion-managed fiber optic system due to the interplay among dispersion, nonlinearity and partial coherence of the source. When the source is fully coherent, the nonlinear mixing of signal pulses generate echo pulses due to intra-channel four wave mixing (IFWM). The echo pulses interfere with the signal pulses leading to large power fluctuations at the fiber output. When the fiber is excited by a partially coherent source, the efficiency of IFWM generation is reduced. In addition, the visibility of interference fringes (in time domain) due to the interference of the echo pulses and signal pulses drops as the coherence time decreases. As a result, the mean nonlinear distortion is reduced as the coherence time decreases in amplitude modulated systems.We have developed an analytical expression for the mean nonlinear distortion due to the interference of a signal pulse with an echo pulse generated due to the nonlinear interaction of signal pulses when the fiber is excited by a partially coherent source. For DU systems, as the spectral width of the source changes from 2 kHz to 2 GHz), the mean nonlinear distortion drops by 14% whereas for DM systems, it hardly changes. Our numerical simulation result is found to be consistent with the analytical predictions.

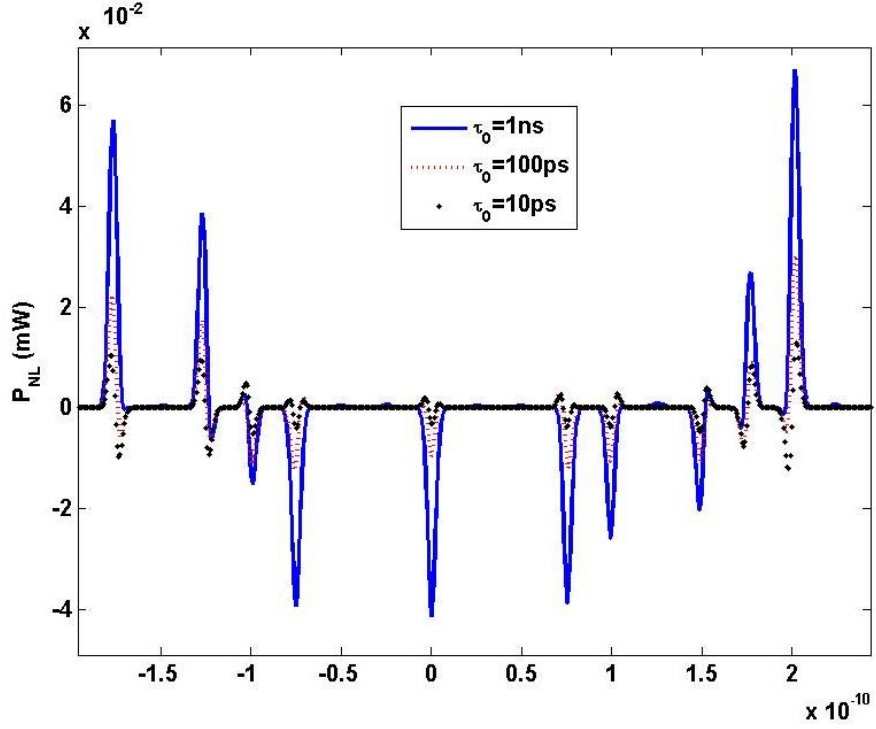

Fig.7. Ensemble-averaged nonlinear distortion as a function of time for various coherence times in DU system.

Table 2: Mean nonlinear distortion for the DU system. $T_w$=640 ps.

| Coherence time $\tau_0$ (ns) | $\sigma_{NL}$ (mW) |
|---|---|
| 1 | 0.0097 |
| 0.1 | 0.0037 |
| 0.01 | 0.0020 |

**Appendix A**

Taking the Fourier transform of Eq. (27), we obtain

$$i\frac{\partial \tilde{\varepsilon}(\omega)}{\partial z}+\frac{\beta_2}{2}\omega^2 \tilde{\varepsilon}(\omega)=-\gamma e^{-\alpha z} F\{u_j u_k u_l^*\}. \tag{39}$$

Plugging Eq. (23) to the RHS of Eq. (39), we find

$$u_j u_k u_l^* = a_j a_k a_l \frac{|A(t)|^2 A(t) T_0^3}{|T_1|^2 T_1} e^{-\frac{(t-jT_s)^2}{2T_1^2}-\frac{(t-kT_s)^2}{2T_1^2}-\frac{(t-lT_s)^2}{2T_1^{*2}}}, \tag{40}$$

$$= a_j a_k a_l \frac{|A(t)|^2 A(t) T_0^3}{|T_1|^2 T_1} e^{-\Lambda t^2 + Bt + C},$$

where

$$\Lambda = \frac{3T_0^2 + iS}{2(T_0^4 + S^2)}, \tag{41}$$

$$B = \frac{T_s\left[(j+k+l)T_0^2 + i(j+k-l)S\right]}{T_0^4 + S^2},\tag{42}$$

and

$$C = -\frac{T_s^2\left[(j^2+k^2+l^2)T_0^2 + i(j^2+k^2-l^2)S\right]}{2(T_0^4 + S^2)}.\tag{43}$$

Substituting Eq. (40) into Eq. (39), and solving the first order differential equation, we find

$$\tilde{\varepsilon}(\omega,z) = a_j a_k a_l i\gamma T_0^3 e^{i\beta_2\omega^2 z/2} \int_0^z ds\, e^{-\alpha s} e^{-i\beta_2\omega^2 s/2}$$
$$\times \int_{-\infty}^{+\infty} \frac{1}{|T_1(s)|^2 T_1(s)} |A(t')|^2 A(t') e^{-At'^2 + Bt'^2 + C} e^{-i\omega t'} dt'.\tag{44}$$

Taking the inverse Fourier transform of Eq. (44), we obtain

$$\varepsilon(t,z) = \frac{i}{2\pi} T_0^3 \gamma a_j a_k a_l \int_{-\infty}^{+\infty}\int_0^z d\omega ds\, e^{-\alpha s + i\beta_2\omega^2(z-s)/2 + i\omega t} \int_{-\infty}^{+\infty} dt' \frac{1}{|T_1(s)|^2 T_1(s)} |A(t')|^2 A(t') e^{-At'^2 + Bt'^2 + C} e^{-i\omega t'}.\tag{45}$$

**Appendix B**

$A(t)$ is a stationary ergodic random process with the correlation function

$$R(\tau) = \langle A(t) A^*(t+\tau)\rangle = P_{in} e^{-\frac{\tau^2}{\tau_0^2}}.\tag{46}$$

Taking the Fourier transform of Eq. (46), we find

$$F\{R(\tau)\} = \int_{-\infty}^{+\infty} P_{in} e^{-\frac{\tau^2}{\tau_0^2}} e^{-i\omega\tau} d\tau = P_{in} \tau_0 \sqrt{\pi} e^{-\frac{\tau_0^2 \omega^2}{4}}.\tag{47}$$

Let $\tilde{A}(\omega)$ be the Fourier transform of $A(t)$.

$$\tilde{A}(\omega) = \int_{-\infty}^{+\infty} A(t) e^{-i\omega t} dt.\tag{48}$$

Suppose

$$I(\omega,t) = \int_{-\infty}^{+\infty} A^*(t+\tau) e^{-i\omega\tau} d\tau,$$
$$= \int_{-\infty}^{+\infty} A^*(u) e^{-i\omega(u-t)} du,\tag{49}$$
$$= A^*(-\omega) e^{i\omega t}.$$

From Eqs. (46) and (47)

$$F\{R(\tau)\} = \int_{-\infty}^{+\infty} d\tau \int_{-\infty}^{+\infty} dt A(t) A^*(t+\tau) e^{-i\omega\tau},$$
$$= \tilde{A}^*(-\omega) \int_{-\infty}^{+\infty} A(t) e^{i\omega t} dt,$$
$$= |\tilde{A}(-\omega)|^2 = P_{in} \tau_0 \sqrt{\pi} e^{-\frac{\tau_0^2 \omega^2}{4}}.\tag{50}$$

Let

$$\tilde{A}(\omega) = N(\omega) e^{i\theta(\omega)},\tag{51}$$

where $N(\omega)$ and $\theta(\omega)$ are amplitude and phase, respectively. From Eq. (50), we find

$$N(\omega) = \sqrt{P_{in}\tau_0}\, \pi^{1/4} e^{-\frac{\tau_0^2 \omega^2}{8}}.\tag{52}$$

Let

$$R_{NL}(\tau) = \left\langle |A(t)|^2 A(t) A(t+\tau)^* \right\rangle = \int_{-\infty}^{+\infty} dt |A(t)|^2 A(t) A^*(t+\tau). \tag{53}$$

Let

$$D(t) = |A(t)^2| A(t) \tag{54}$$

Taking the Fourier transform of Eq. (53), we obtain

$$\begin{aligned} F\{R_{NL}(\tau)\} = \tilde{R}_{NL}(\omega) &= \int_{-\infty}^{+\infty} d\tau \int_{-\infty}^{+\infty} dt |A(t)|^2 A(t) A^*(t+\tau) e^{-i\omega\tau}, \\ &= \int_{-\infty}^{+\infty} I(\omega,t) D(t) d\tau, \\ &= \int_{-\infty}^{+\infty} D(t) \tilde{A}^*(-\omega) e^{i\omega t} d\tau, \\ &= \tilde{A}^*(-\omega) \tilde{D}(-\omega). \end{aligned} \tag{55}$$

Suppose $G(t) = |A(t)^2|$. Fourier transforming Eq. (54), we find

$$\begin{aligned} \tilde{D}(\omega) &= F\{G(t)A(t)\}, \\ &= \frac{1}{2\pi} \int_{-\infty}^{+\infty} \tilde{G}(\omega-\Omega_2) \tilde{A}(\Omega_2) d\Omega_2, \end{aligned} \tag{56}$$

where

$$\tilde{G}(\omega) = F\{G(t)\} = \frac{1}{2\pi} \int_{-\infty}^{+\infty} \tilde{A}(\omega-\Omega_1) \tilde{A}^*(-\Omega_1) d\Omega_1. \tag{57}$$

So, Eq. (56) becomes

$$\tilde{D}(\omega) = \frac{1}{4\pi^2} \int_{-\infty}^{+\infty} \int_{-\infty}^{+\infty} \tilde{A}(\omega-\Omega_2-\Omega_1) \tilde{A}^*(-\Omega_1) \tilde{A}(\Omega_2) d\Omega_1 d\Omega_2. \tag{58}$$

From Eq. (55), we have

$$\begin{aligned} \tilde{R}_{NL}(\omega) &= \tilde{A}^*(-\omega) \tilde{D}(-\omega), \\ &= \frac{1}{4\pi^2} \sqrt{P_{in}\tau_0} \pi^{1/4} e^{-\frac{\tau_0^2 \omega^2}{8} - i\theta(-\omega)} \int_{-\infty}^{+\infty} \int_{-\infty}^{+\infty} \tilde{A}(-\omega-\Omega_2-\Omega_1) \tilde{A}^*(-\Omega_1) \tilde{A}(\Omega_2) d\Omega_1 d\Omega_2, \\ &= \frac{1}{4\pi} P_{in}^2 \tau_0^2 e^{-\frac{\tau_0^2\omega^2}{8} - i\theta(-\omega)} \int_{-\infty}^{+\infty} \int_{-\infty}^{+\infty} e^{-\frac{\tau_0^2(-\omega-\Omega_2-\Omega_1)^2}{8} - \frac{\tau_0^2 \Omega_1^2}{8} - \frac{\tau_0^2 \Omega_2^2}{8}} e^{i\theta(-\omega-\Omega_2-\Omega_1) - i\theta(-\Omega_1) + i\theta(\Omega_2)} d\Omega_1 d\Omega_2. \end{aligned} \tag{59}$$

Let

$$\begin{aligned} \Theta(\omega) &= \left\langle e^{-i\theta(-\omega) + i\theta(-\omega-\Omega_2-\Omega_1) - i\theta(-\Omega_1) + i\theta(\Omega_2)} \right\rangle, \\ &= 1 \text{ if } \Omega_1 = -\Omega_2 \text{ or } \omega = -\Omega_2, \\ &= 0 \text{ otherwise.} \end{aligned} \tag{60}$$

Substituting Eq. (60) in Eq. (59), we find

$$\tilde{R}_{NL}(\omega) = \frac{1}{2\sqrt{\pi}} P_{in}^2 \tau_0 e^{-\frac{\tau_0^2\omega^2}{4}}. \tag{61}$$

Inverse Fourier transforming Eq. (61), we obtain

$$R_{NL}(\tau) = \left\langle A(t) |A(t)|^2 A(t+\tau) \right\rangle = \frac{1}{2\pi} P_{in}^2 e^{-\frac{\tau^2}{\tau_0^2}}. \tag{62}$$